\documentclass[aps,prd,a4paper,superscriptaddress,preprint,nofootinbib]{revtex4-1}
\pdfoutput=1
\usepackage{amsfonts}
\usepackage{amsmath}
\usepackage{hyperref}
\usepackage{graphicx}

\def\321{$SU(3)\times SU(2)\times U(1)$}

\def\hmu{\hat{\mu}}

\def\10{SO(10)}

\hypersetup{
pdfstartview=FitV,
colorlinks=true, 
linkcolor=blue, 
citecolor=red, 
filecolor=black, 
urlcolor=blue}

\newcommand{\tb}{\tan\beta}
\newcommand{\stopr}{\tilde{t}_R}
\newcommand{\stopl}{\tilde{t}_L}
\newcommand{\hu}{H_u}
\newcommand{\hd}{H_d}
\newcommand{\msq}[1]{m^2_{#1}}
\newcommand{\lb}{\left(}
\newcommand{\rb}{\right)}
\newcommand{\yuk}[1]{y_{#1}}
\newcommand{\vcal}{\mathcal{V}}

\newcommand{\ms}[2]{m_{\tilde{#1}_{#2}}}
\newcommand{\msusy}{M_{\rm SUSY}}
\newcommand{\MW}{M_{W}}

\newcommand{\gev}{\rm GeV}

\newcommand{\staur}{\tilde{\tau}_R}
\newcommand{\staul}{\tilde{\tau}_L}
\newcommand{\rsix}{\sqrt{6}}

\begin{document}
\title{Charge and Color Breaking Constraints in MSSM after the Higgs Discovery at LHC}

\author{Debtosh Chowdhury}
\email{debtosh.chowdhury@roma1.infn.it}
\affiliation{Istituto Nazionale di Fisica Nucleare, Sezione di Roma, Piazzale Aldo Moro 2, I-00185, Rome, Italy}
\affiliation{Centre for High Energy Physics, Indian Institute of Science, Bangalore 560 012, India}
\author{Rohini M. Godbole}
\email{rohini@cts.iisc.ernet.in}
\author{Kirtimaan A. Mohan}
\email{kirtimaan@cts.iisc.ernet.in}
\author{Sudhir K. Vempati}
\email{vempati@cts.iisc.ernet.in}
\affiliation{Centre for High Energy Physics, Indian Institute of Science, Bangalore 560 012, India}

\date{\today}

\begin{abstract}
\vskip 0.7cm
We revisit the constraints on the parameter space of the Minimal Supersymmetric Standard Model (MSSM), from charge and color breaking minima in the light of information on the Higgs from the LHC so far. We study the behavior of the scalar potential keeping two light sfermion fields along with the Higgs in the pMSSM framework and analyze the stability of the vacuum. We find that for lightest stop $\lesssim 1$ TeV and small $\mu \lesssim 500$ GeV , the absolute stability of the potential can be attained only for  $|X_{t}| \lesssim  \sqrt{ 6 m_{\tilde{t}_1} m_{\tilde{t}_2}}$. The bounds become stronger for larger values of the $\mu$ parameter. Note that this is approximately the value of $X_t$ which maximizes the Higgs mass. Our bounds on the low scale MSSM parameters are more stringent than those reported earlier in literature. We reanalyze the stau sector as well, keeping both staus. We study the connections between the observed Higgs rates and vacuum (meta)stability. We show how a precision study of the ratio of signal strengths, ($\mu_{\gamma\gamma}/\mu_{ZZ}$) can shed further light. 
\end{abstract}

\maketitle


\section{Introduction}

The 7 and 8 TeV runs of the large hadron collider, the LHC, seem to have provided us with evidence for the last missing piece of the Standard Model(SM) : a new boson with properties very similar to those expected of the SM Higgs boson \cite{Chatrchyan:2012ufa,Aad:2012tfa,CMS-PAS-HIG-13-005,ATLAS-CONF-2013-034}. The low mass of the observed state is completely consistent with the indirect constraints on the Higgs mass in the SM implied by the electroweak precision measurements \cite{Baak:2012kk,Baak:2011ze}.
New physics beyond the SM (BSM) proposed to stabilize the Higgs mass close to the electroweak scale was expected to reveal itself at the LHC. 
However, the present runs of the LHC have not yielded any evidence for the same and have only resulted in  bounds on the  masses of the new expected particles in various BSM scenarios.  

Along with stabilization of the Higgs mass around the electroweak scale against radiative corrections, there are quite a few other pointers 
towards BSM physics such as Dark Matter (DM), observed baryon asymmetry in the Universe (BAU), neutrino masses, a lack of explanation of the large mass hierarchy observed in the fermion mass spectrum etc. However, the former is the only motivation  that necessarily predicts BSM physics around the TeV scale and hence is under significant pressure in light of the lack of BSM signals at the LHC. At the same time, the close connection with the Higgs sector that any such BSM physics has, implies that a precision study of the Higgs sector is sure to provide us with a good probe of the same.

Just the mass of the observed Higgs state can give us a lot of information about both the SM and the BSM. The experimentally observed mass of the state is an extremely interesting value. In the SM, it is just large enough to indicate possible instability of the vacuum  at very high scales, which crucially depends on the exact values of $m_h, m_t$ and $\alpha_s$ \cite{Cabibbo:1979ay,Ellis:2009tp,EliasMiro:2011aa,EliasMiro:2012ay,Degrassi:2012ry,Zoller:2012cv}. Thus the observed mass of the Higgs may be incompatible with absolute stability of the vacuum. However,  there is a possibility that the electroweak vacuum can  end up being meta-stable i.e., its life time could be larger than the age of the universe and no new physics below the Planck scale may be necessary to stabilize the SM vacuum\footnote{ For a discussion on the effect of Planck scale dynamics on destabilizing the electroweak vacuum see  for example Ref.~\cite{Branchina:2013jra}. } \cite{Isidori:susy13}.

If the BSM physics in question is Supersymmetry the mass of the lightest Higgs state is in fact bounded. The observed mass ($\sim 125 $ GeV) is very close to (and smaller than) the upper bound expected in the simple supersymmetric extension of the SM (the MSSM) and already constrains the relevant parameters of the MSSM rather strongly.\footnote{It is also interesting to note that the closeness of the observed Higgs mass to the upper bound implies that the sparticle masses be around the TeV scale and thus in fact quite consistent with the non observation of the direct SUSY signal so far.} 
If the Higgs boson were observed with a mass above the bound, this could have easily ruled out MSSM. 

In case of  Supersymmetry with its extended scalar sector, the appearance of the additional minima of the scalar potential is a natural feature. These minima can 
be either color or charge breaking (CCB). The scalar potential of the MSSM is more constrained than in the SM as some of the quartic couplings are simply related to the gauge couplings. The relative ordering of the different minima depends on the parameters of the Supersymmetric theories such as the sparticle masses and the trilinear couplings. The multiple minima of the scalar potential of the MSSM have been studied in literature in great detail  \cite{Frere:1983ag,AlvarezGaume:1983gj,Derendinger:1983bz,Kounnas:1983td,Drees:1985ie,Gunion:1987qv,Komatsu:1988mt,Langacker:1994bc,Casas:1995pd}. 

In the present work we revisit the charge and color breaking minima in the light of the discovery of the Higgs mass and the results from LHC. We study the CCB bounds in the presence of light stops and light staus. We find bounds stronger than those presented earlier in literature~\cite{Kusenko:1996jn}. 

The rest of the paper is organized as follows.
In the next section we outline the connection between the Higgs mass and stability of the vacuum in the MSSM. In section~\ref{sec:CCB} we describe the MSSM potential and discuss the results for the case of light stops. In section~\ref{sec:staus} we discuss the implications of light staus on stability and Higgs decays. We summarize our results and conclude in section~\ref{sec:summary}.


\section{Stability and light Higgs mass in MSSM}
\label{sec:mh-stab}

The recent discovery of the Higgs boson at the LHC  around 125 GeV \cite{Chatrchyan:2012ufa,Aad:2012tfa} puts severe constraints on the MSSM \cite{Hall:2011aa,Heinemeyer:2011aa,AlbornozVasquez:2011aa,Arbey:2011aa,Arbey:2011ab,Draper:2011aa,Carena:2011aa,Christensen:2012ei,Baer:2011ab,Kadastik:2011aa,Buchmueller:2011ab,Aparicio:2012iw,Ellis:2012aa,Baer:2012uya,Arbey:2012dq,Cao:2012fz}. The present results put the measured mass to lie within  \cite{ATLAS-CONF-2013-014,CMS-PAS-HIG-13-005}  
\begin{align}
m_h = 125.7 \pm 0.6\ \text{GeV}\ .  
\end{align}
While such a value of the Higgs mass can be accommodated within MSSM, this would require a special choice of the parameters, in particular the trilinear stop mixing term, $X_t$.  Including the leading one loop corrections, the lightest CP even Higgs in the MSSM has a mass given by \cite{Haber:1990aw,Ellis:1990nz,Okada:1990vk,Haber:1996fp}
\begin{align}
m_h^2 \approx   M_Z^2 \cos^2 2\beta\ +&\ \frac{3 g_2^2 m_t^4}{8 \pi^2 \MW^2} \left[\ln\left(\frac{\ms{t}{1} \ms{t}{2}}{m_t^2}\right) + \frac{X_t^2}{\ms{t}{1}\ms{t}{2}} 
\left(1 - \frac{X_t^2}{12\ms{t}{1}\ms{t}{2}} \right) \right]\ ,
\label{higgs1l}
\end{align}
where $X_t = A_t - \mu/\tan \beta$. The first term is the usual tree level mass term and the second term is the dominant
1-loop correction from the top-stop loop.  For stops of the $\mathcal{O}(1$ TeV), a value of 125 GeV for the mass of the light Higgs requires a significant contribution from the $X_t$ terms. The contribution from these terms gets maximized for values
of $|X_t| \sim \sqrt{6 ~m_{\tilde{t}_1} m_{\tilde{t}_2}}$.  Such large values for $X_t$  typically translate into 
large values of the trilinear coupling $A_t ~\sim 1$ TeV, i.e. comparable to the stop masses. 
\begin{figure}
\centering
\includegraphics[width=0.48\textwidth]{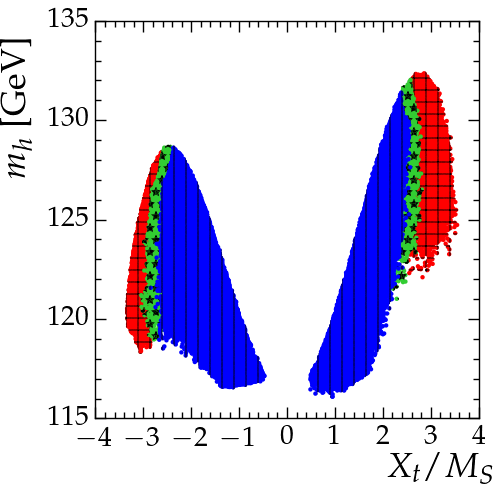}
\includegraphics[width=0.48\textwidth]{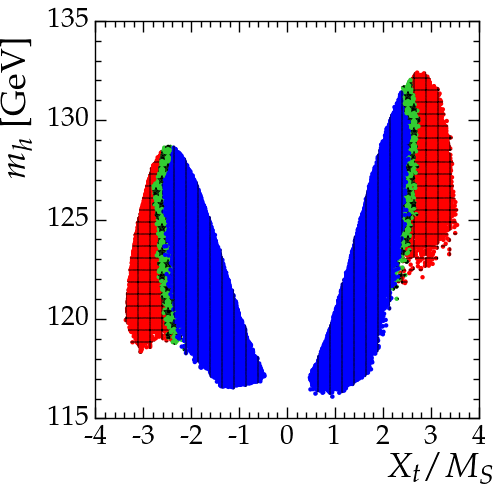}
\caption{Stable (blue, vertical lines), meta-stable (green, stars) and unstable (red, checkered) vacuum in the $m_h$ vs. $X_t/M_S$  plane.  The left panel represents three field analysis and the right panel four field analysis.}
\label{higgsvacuum}
\end{figure}
It has been known that large values of $A_t$
can lead to charge and color breaking minima \cite{Frere:1983ag,AlvarezGaume:1983gj,Derendinger:1983bz,Kounnas:1983td,Drees:1985ie,Gunion:1987qv,Komatsu:1988mt,Langacker:1994bc,Casas:1995pd}\footnote{Alternate sources of CCB have also been suggested as in Ref.~\cite{Cline:1999wi}.}.
In light of this, it would be interesting to know whether large $A_t$ values required to satisfy the Higgs mass measurement  lead to color and charged breaking (CCB) minima. We have done an exhaustive numerical analysis (as will be detailed in the next section) looking for charge and color
breaking minima in the field space of the two Higgs doublets and $\tilde{t}_L$ and $\tilde{t}_R$.  We have classified the various possible minima as follows:
\begin{itemize}
\item \textit{Unstable Minima:} In this case the minima are deeper than the electroweak minima in the stop field space and
the electroweak vacuum will have a life time less  than the age of the Universe and thus unstable. 
\item  \textit{Meta-stable Minima:}  In this case, even though there are deeper CCB minima compared to the electroweak minima, 
the decay time is greater than the age of the universe and thus the Universe resides in a meta-stable vacuum. 
\item \textit{Stable Minima:} There are no deeper minima in the vicinity of the electroweak vacuum which is now stable. We add that the EWSB minimum is only stable in the subset of field space that we consider and inclusion of other field directions could destabilize the EWSB minimum. 
\end{itemize}

In the plane of the Higgs mass and $X_t/M_S$, we have plotted the regions of the three kinds of vacuum in 
Fig.~\ref{higgsvacuum}. $M_S$ $(= \sqrt{m_{\tilde{t}_1} m_{\tilde{t}_2}})$ represents the typical scale of soft Supersymmetry breaking parameters. 

The Higgs mass is roughly given by Eq.(\ref{higgs1l}),while for numerical analysis presented in the figure we use the  full two-loop Higgs mass formula~\cite{Chowdhury:2011zr}. The left panel presents the results from a three field analysis (up type Higgs ($H_u$),
$\tilde{t}_L$, $\tilde{t}_R$ ), whereas the right panel presents the results from a four field analysis ($H_u$, down type Higgs  ($H_d$), $\tilde{t}_L$, $\tilde{t}_R$). We have assumed pMSSM like boundary conditions~\cite{Djouadi:1998di} which has 22 input parameters. Out of the twenty-two parameters  the most relevant parameters for the analysis are just the stop and Higgs sector parameters ($m_A,\tb,m_{\tilde{Q}_{33}},m_{\tilde{U}_{33}},\mu$).

From Fig.~\ref{higgsvacuum} we see that  CCB constraints start playing an important role for $|X_t| \gtrsim \rsix M_S$. We define $|X_t^{\text{max}}| \equiv \sqrt{6}M_S$, when its contribution to the light Higgs mass is the maximum. Beyond this value, there is a thin band of points where the electroweak vacuum is meta-stable, and further down unstable. Given that we consider pMSSM boundary conditions, these results are fairly general and apply for a wide class of SUSY breaking models. 
The results also show that for lightest stop $\lesssim \mathcal{O}(1$ TeV) a Higgs mass of $\sim 125$ GeV implies that
there exists a value of $X_t$ close to its
critical value, $X_t^{\text{max}}$, beyond 
which the vacuum is either meta-stable or unstable.  From Fig.~\ref{higgsvacuum}, we see that there is hardly any difference in the 
vacuum stability regions between the three field and four field analysis. This is only true as long as the $\mu$ parameter takes values
less than the stop masses. Furthermore, as we will see in the next section, if $\mu$ is much larger than the stop masses, 
 the vacuum starts becoming meta-stable for smaller values of $X_t $, $|X_t|  < |X_t^{\text{max}}|$. This is an important point to note. 
Finally,  for larger stop masses ($\gtrsim 4$ TeV) the condition of meta-stability and stability would still be $|X_t| \lesssim \rsix M_S$. 
However the Higgs mass $\sim 125$ GeV does not require such large $X_t$ values.

\section{Revisiting CCB in MSSM}
\label{sec:CCB}

The scalar potential of the MSSM consists of all the super-partners of the SM fermions. Electroweak symmetry breaking (EWSB) requires that the Higgs field should acquire a nonzero vacuum expectation value (vev). The potential at this value of the Higgs fields ($v_u$, $v_d$) corresponds to a minimum of the potential (EWSB minimum). However the complete potential may have other minima in the field directions where charged and/or colored particles acquire nonzero vevs. This corresponds to the existence of a charge and/or color breaking minimum of the potential. In such cases it is possible that the EWSB minimum may not be the global minimum of the potential. It may well be that our universe sits in this local minimum of EWSB, thus making it a false vacuum. Transition from the local EWSB minimum to the global CCB minimum happens through quantum tunneling and is inevitable. As long as the time associated with the  transition from the local EWSB minimum to the deeper CCB minimum is greater than the age of the universe, this is an acceptable configuration of the potential.

The probability of decay per unit time per unit volume ($\Gamma/V$) of the false or meta-stable vacuum to a deeper vacuum can be estimated using semi-classical techniques developed in \cite{Coleman:1977py,Callan:1977pt}. In the semi-classical limit this quantity is given by
\begin{align}\label{eqn:trans}
\frac{\Gamma}{V} = A\, e^{-S[\bar{\phi}]/\hbar}\ ,
\end{align} 
where $V$ corresponds to the volume, $S[\bar{\phi}]$ is the Euclidean action evaluated on the bounce configuration and $\Gamma$ is the width associated with the tunneling of a particle from the false vacuum to the deeper vacuum. The prefactor $A$ is roughly of the order of the fourth power of the scale associated with the potential and we set it to $(100\ \gev)^4$ \cite{Kusenko:1996jn}. The condition that the lifetime of the false vacuum be larger than the age of the universe implies that $\Gamma/V$ must be smaller than the fourth power of the Hubble constant ($H_0 \sim 1.44 \times 10^{-42}\ \gev$). This can be translated to a condition on the value of the action such that $S[\bar{\phi}]/\hbar \gtrsim 404$ \cite{Claudson:1983et,Kusenko:1996jn}.

The scalar potential of the MSSM consists of squared, cubic and quartic terms of the various scalar fields and is therefore quite complicated. For a physically viable spectrum, the bi-linear terms (quadratic terms of the potential) must be positive\footnote{The Higgs sector is an exception where the bi linear term proportional to $B_{\mu}$ is negative which is required for symmetry breaking.} while the quartic terms always contribute positively to the potential. The trilinear terms which can be negative are therefore responsible for the formation of additional minima other than the EWSB minimum\footnote{ We note that large values of $A_t$ can have other interesting consequences as suggested in Ref.~\cite{Pearce:2013yja}.}. If the bilinear terms are large then they tend to mitigate the destabilizing effect of the trilinear terms. One therefore does not have to consider the full MSSM potential but rather only the fields that have lighter masses and large trilinear couplings. For the case of stops the trilinear terms are also enhanced by the relatively large value of the Yukawa couplings. 

We now proceed to a description of the potential and a discussion of the results in detail. For simplicity let us first consider the potential consisting only of $(\hu,\stopl,\stopr)$.

\subsection{Three field scalar potential in MSSM}
\label{sec:CCB-3f}

The tree-level scalar potential in MSSM in the $\hu, \stopl$ and $\stopr$ field directions is
\begin{align}
\mathcal{V}_3 = & \lb \msq{\hu} + \mu^2 \rb \left|\hu\right|^2 + \msq{\stopl} | \stopl |^2 + \msq{\stopr} \left| \stopr \right|^2 + \lb \yuk{t} A_t\,  \hu^* \stopl \stopr + {\rm c.c.} \rb \notag \\ 
&\ + \yuk{t}^2 \lb | \stopl \stopr |^2 + | \hu \stopl |^2 + | \hu \stopr |^2 \rb + \frac{g_1^2}{8} \lb  |\hu|^2 + \frac{1}{3} |\stopl|^2 - \frac{4}{3} |\stopr|^2  \rb^2 \notag \\
&\ + \frac{g_2^2}{8} \lb |\hu|^2 - |\stopl|^2 \rb^2 + \frac{g_3^2}{6} \lb |\stopl|^2 - |\stopr|^2 \rb^2\ .
\label{eqn:3f}
\end{align}
The first observation that one can make is that this potential is not unbounded from below (UFB) in any of the field directions. This is due to the $F$-terms that contribute positively to the potential for nonzero values of the field. The second observation is, as mentioned earlier, one can always choose the phases of the fields in such a way that the trilinear term $\lb \yuk{t} A_t\,  \hu^* \stopl \stopr + {\rm c.c.} \rb$ contributes negatively to the potential. Thus it is the only term which is responsible for the formation of new CCB minima.

Before we turn to a discussion of the numerical analysis we first give a qualitative understanding of the development of CCB minima. In the $D$-flat direction the $D$-terms, which are fourth power in fields and  hence always positive, will contribute minimally to the potential. It is therefore likely that CCB minima exist in the $D$-flat direction.
This direction corresponds to the 
choice where $|\hu| = |\stopl| = |\stopr| = \xi$. With this  choice of field directions, Eq.(\ref{eqn:3f}) reduces to 
\begin{align}\label{eqn:3fs}
\mathcal{V}_3 = & \lb \msq{\hu} + \mu^2  +\msq{\stopl} + \msq{\stopr} \rb \xi^2  - 2 \yuk{t} |A_t|\, \xi^3 + 3\yuk{t}^2 \xi^4 \ .
\end{align}
Taking the derivative of Eq.(\ref{eqn:3fs}) with respect to $\xi$ and solving one finds that the condition for existence of a CCB minimum is $A_t^2 \gtrsim 2.67\lb \msq{\hu} + \mu^2  +\msq{\stopl} + \msq{\stopr} \rb  $. However, it has been shown that the existence of a CCB minimum \textit{deeper} than the EWSB minimum corresponds to ~\cite{Claudson:1983et}
\begin{align}\label{eqn:stab}
A_t^2 > 3\lb \msq{\hu} + \mu^2  +\msq{\stopl} + \msq{\stopr} \rb\ .
\end{align}


We now proceed to discuss the CCB bound in MSSM through detailed numerical analysis. It is well known that the twenty-two parameter phenomenological MSSM (pMSSM) provides the most economical set to describe low energy SUSY phenomenology without considering any particular SUSY breaking scenario at the high scale. For the purpose of studying light stops it is not necessary to vary all twenty-two parameters. We therefore vary the stop sector and fix all other sectors of MSSM. More specifically our choice of parameters are 
\begin{align}\label{scan-ranges}
\tan\beta &\in [5,\ 60] \notag \\ 
m_{\tilde{Q}_{33}} & \in [500,1500]\ \gev \notag \\
m_{\tilde{U}_{33}} & \in [500,1500]\ \gev \\
\mu &\in [100,1000]\ \gev \notag \\
A_t &\in [-3,\ 3]\, m_{\tilde{Q}_{33}} \notag .
\end{align}
We set all other sfermion masses to be at $1000$ GeV, $M_A=1000$ GeV and all other trilinear couplings are set to zero. The gaugino masses are set at $M_{1} = 100$ GeV, $M_2  = 300$ GeV and $M_3 = 1000$ GeV. We note that this choice of parameters corresponds to the decoupling regime of the MSSM Higgs sector~\cite{Djouadi:rep2}. We impose on the resulting spectrum, the flavor constraints given below.
\begin{align}
\label{indirect-limits}
{\rm BR}(B \rightarrow X_s \gamma) &\in [2.99,\ 3.87]\times 10^{-4}\  \textrm{\cite{Amhis:2012bh}}, \nonumber \\
{\rm BR}(B_s \rightarrow \mu^+ \mu^-) &\in [1.2,\ 5.0]\times 10^{-9}\ \textrm{\cite{Aaij:2013aka,*Chatrchyan:2013bka}}. 
\end{align}
Additionally we demand that the lightest stop mass lies in the range $[500,1000]$ GeV.
We generate the SUSY spectrum with a modified version of {\tt SuSeFLAV} \cite{Chowdhury:2011zr}, the Higgs spectrum is calculated using \texttt{FeynHiggs 2.10.4} \cite{Heinemeyer:1998yj,Heinemeyer:1998np,Degrassi:2002fi,Frank:2006yh,Heinemeyer:2007aq}, the flavor constraints are calculated using {\tt micrOMEGAs 3.2} \cite{Belanger:2013oya} and in order to calculate the transition rate of the decay  of the false vacuum at zero temperature we use {\tt CosmoTransitions} \cite{Wainwright:2011kj}.

\begin{figure}
\centering
\includegraphics[width=0.48\textwidth]{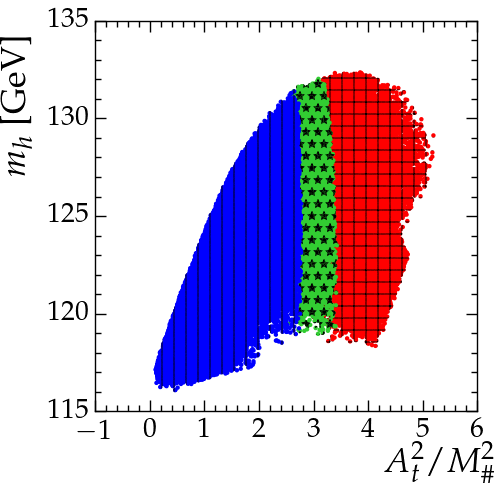}
\includegraphics[width=0.48\textwidth]{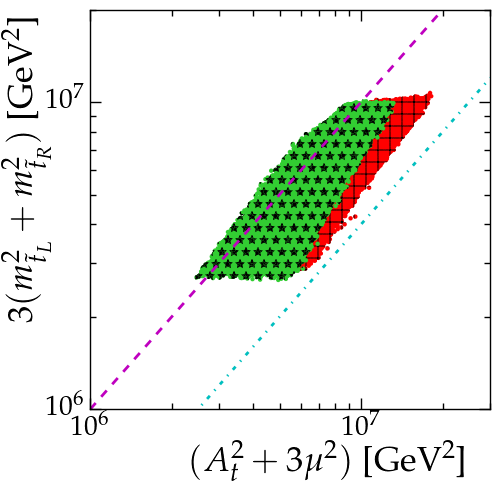}
\caption{Left: The variation of $m_h$ against $A_t^2/M_\#^2$ for three field potential. 
Right: The variation of $3(\msq{\stopl} + \msq{\stopr})$ with $A_t^2 + 3\mu^2$ for the three field. The dashed (magenta) line corresponds to the analytic bound and the dot-dashed (cyan) line corresponds to the empirical bound. Points that correspond to the EWSB vacuum being unstable are given in red (checkered), meta-stable in green (stars) and stable in blue (vertical lines). }
\label{fig:mhash-3f}
\end{figure}


The left plot of Fig.~\ref{fig:mhash-3f} shows the lightest Higgs mass against $A_t^2/M_\#^2$ where $M_\#^2=\lb \msq{\hu} + \mu^2  +\msq{\stopl} + \msq{\stopr} \rb$. Points in blue correspond to points of absolute stability, i.e. where the EWSB minimum is a global minimum. Points in red and green correspond to a CCB minimum being the global minimum; the green points represent those points of the parameter space where the time associated with the transition from the EWSB minimum to the CCB minimum is greater than the age of the universe(meta-stable) and red points are those where this time is less than the age of the universe(unstable).
We see from  Fig.~\ref{fig:mhash-3f}, the region of absolute stability is slightly diminished in comparison to the bound given in Eq.(\ref{eqn:stab}).
The result quoted earlier (derived by looking at the most likely direction of formation of CCB minima ($D$-flat direction)) is only an approximate result since we effectively reduce the 3 dimensional problem to a one dimensional problem. We also note that the regions of meta-stability do not extend to very large values of $A_t$. We therefore find stronger bounds than the empirical bound found in Ref.~\cite{Kusenko:1996jn} for the constrained MSSM (cMSSM). To emphasize this point we show in the right hand side of Fig.~\ref{fig:mhash-3f} a plot similar to the one found in Ref.~\cite{Kusenko:1996jn}. The dashed (magenta) line in this plot corresponds to the analytic bound 
\begin{align}
A_t^2 + 3\mu^2 < 3(\msq{\stopl} + \msq{\stopr})\ ,
\end{align}
and the dot-dashed (cyan) line corresponds to the empirical bound 
\begin{align}
A_t^2 + 3\mu^2 < 7.5(\msq{\stopl} + \msq{\stopr})\ .
\end{align}
We see that regions of meta-stability do not extend to the empirical bound.

Our results do not have a strong dependence on $\tb$. It can be shown that for moderately large $\tb$, the $\tb$ dependence of the coefficients in the potential factor out. Note that the potential $\vcal_3$ has no $\mu$ dependence. We further checked that the results are insensitive to the variation of $m_A$ between $500-2000$ GeV. We now proceed to describe our results when the fourth $\hd$ field is added to the potential.


\subsection{Four field scalar potential in MSSM}
\label{sec:CCB-4f}

\begin{figure}
\centering
\includegraphics[width=0.48\textwidth]{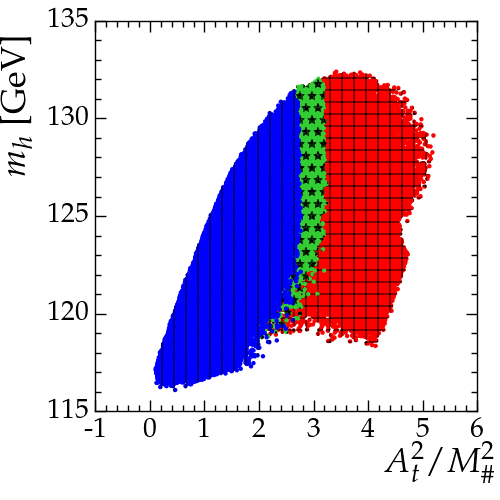}
\includegraphics[width=0.48\textwidth]{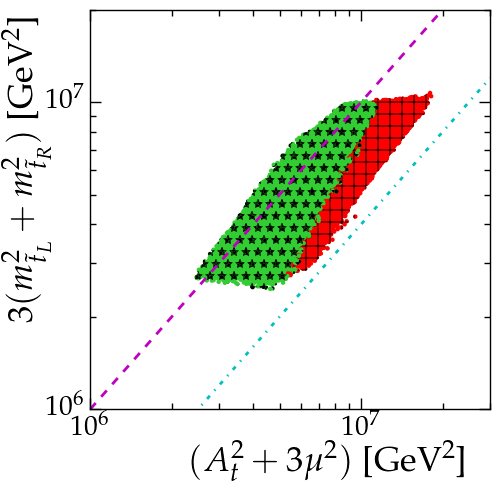}
\caption{Left: The variation of $m_h$ against $A_t^2/M_\#^2$ for the four field potential. 
Right: The variation of $3(\msq{\stopl} + \msq{\stopr})$ with $A_t^2 + 3\mu^2$ for the four field potential. The dashed (magenta) line corresponds to the analytic bound and the dot-dashed (cyan) line corresponds to the empirical bound. Points that correspond to the EWSB vacuum being unstable are given in red (checkered), meta-stable in green (stars) and stable in blue (vertical lines).}
\label{fig:mhash-4f}
\end{figure}

The tree-level potential in MSSM due to the presence of $\stopl$ and $\stopr$ field can be expressed as 
\begin{align}
\vcal_{4} = &\; \lb \msq{\hu} + \mu^2 \rb |\hu|^2 + \lb \msq{\hd} + \mu^2 \rb |\hd|^2 + 
\msq{\stopl} |\stopl|^2 + \msq{\stopr} |\stopr|^2 - \notag \\ 
&\; B_{\mu} \lb \hu \hd + {\rm c.c.} \rb  + \lb \yuk{t} A_{t} \hu \stopl \stopr +  {\rm c.c.} \rb - \lb \yuk{t} \mu \stopl \stopr \hd^{*} + {\rm c.c.} \rb + \notag \\
&\; \yuk{t}^2 \lb | \stopl \stopr |^2 + | \hu \stopl |^2 + | \hu \stopr |^2 \rb + 
\frac{g_2^2}{8} \lb |\hu|^2 - |\hd|^2 - |\stopl|^2 \rb^2 + \notag \\
&\; \frac{g_1^2}{8} \lb |\hu|^2 - |\hd|^2 + \frac{1}{3} |\stopl|^2 - \frac{4}{3} |\stopr|^2 \rb^2 + \frac{g_3^2}{6} \lb |\stopl|^2 - |\stopr|^2 \rb^2\ .
\end{align}

The four-field potential written above, like the three-field potential $\vcal_3$, has no UFB directions. Once again it is easy to see that the most likely direction to find a minimum is in the direction of $|\stopl|=|\stopr|$. It is however not so straight forward to see the other field directions in which a CCB minimum is likely to be present. One must therefore resort to a numerical analysis of the potential. The task is simplified by the fact that one need only look for a minimum within the hypersphere (in the field space) of radius  approximately equal to $A_t + \mu$. For values of the field larger than this value, the quartic terms start dominating, increasing the value of the potential.

\begin{figure}
\centering
\includegraphics[width=0.48\textwidth]{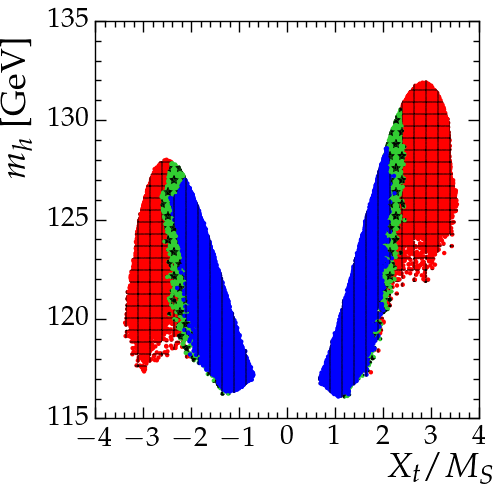}
\includegraphics[width=0.48\textwidth]{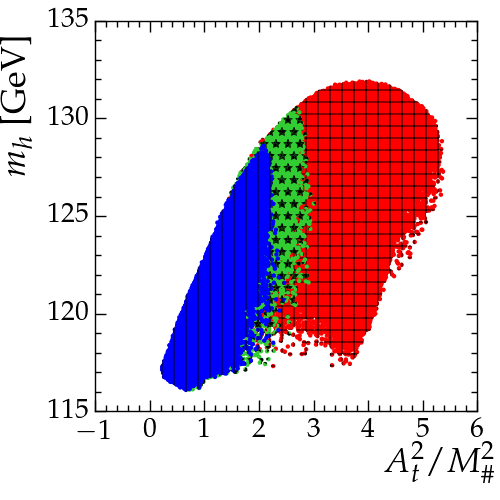}
\caption{Left: The variation of $m_h$ against $X_t$ for the four field potential with $\mu$ set in the range $[1000,1500]$ GeV. Right: The variation of $m_h$ against $A_t^2/M_\#^2$ for the same.  Points that correspond to the EWSB vacuum being unstable are given in red (checkered), meta-stable in green (stars) and stable in blue (vertical lines).}
\label{fig:mhash-4f-mu}
\end{figure}

In Fig.~\ref{fig:mhash-4f} we show the same kind of plots as we did in the three field analysis. We note that our results are very similar to those obtained in the three field analysis.
The presence of the additional $\yuk{t} \mu \stopl \stopr \hd$ term adds negatively to the potential and for large values of $\mu$ unstable points can be found even for small values of $A_t$. We have checked that as one increases the value of $\mu$ the region of absolute stability is further reduced in these plots. This has been shown in Fig.~\ref{fig:mhash-4f-mu}. The left plot shows the variation of $X_t/M_S$ with the Higgs mass whereas the plot on the right shows the variation of $A_t^2/M_\#^2$ with the Higgs mass. We see that in both cases the regions of absolute stability of the vacuum have reduced. The reduction for $X_t/M_S$ is smaller in comparison to $A_t^2/M_\#^2$ which has a significant reduction. Such a feature is absent when one uses the three field potential which is independent of $\mu$. We also check that as expected there is no strong $\tb$ dependence of our results.

The above results have been computed using the tree-level three/four field potentials. 
Radiative corrections to the potential can be important. These can be added in terms
of one loop effective potential corrections. However, choosing a suitable renormalization
scale, the corrections from the one loop effective potential can be minimized. 
If the Higgs fields are the only fields which acquire vevs, choosing the renormalization scale 
close to the weak scale or $\msusy$ would minimize the corrections from the one loop effective potential.

In the presence other sfermionic fields like squarks and sleptons, especially for  
fields with small Yukawa couplings, this choice of the renormalization scale might 
not be appropriate \cite{Gamberini:1989jw}.\footnote{We thank Fabio Zwirner for pointing this out.} 
Given that CCB minima typically appear at large field values
(for the squark/slepton fields $\langle \phi \rangle \lesssim A/2y$, where $A$ is the trilinear coupling and $y$ is the Yukawa coupling), one loop corrections could become
important even if the renormalization scale is chosen to be $\msusy$.  
For stops with the large $\yuk{t}$ appearing inversely in the field values, the
CCB minima are not very far from the weak scale. We have explicitly checked
that they do not exceed 3 TeV for the parameter space we have scanned for the stops.
Hence we do not expect a significant modification of our results presented here by adding one loop corrections. In an upcoming work \cite{ourselves}, we study the renormalization scale dependence of the above bounds in pMSSM in detail.

\section{Light staus and Higgs to Diphoton Decays}
\label{sec:staus}

At the LHC both the production in gluon fusion and the decay rates for $h \to \gamma \gamma$ and $h\to Z\gamma $ final states ate determined by loop induced couplings, which can be affected by the presence of BSM physics. 
Hence, the signal strengths in almost all final states carry an imprint of the same. While the current information from the LHC on signal strengths is consistent with the expectations of the SM Higgs, contributions from BSM physics can not be ruled out. It is interesting to probe the connection between the observed signal strengths of the Higgs and the 
vacuum (meta)stability in the presence of light sfermions. 

The signal strength ($\hmu$) is defined as: 

\begin{equation}
\hmu^{exp}_{XX}=\frac{n^{exp}_{XX}}{\epsilon \sigma(pp \to h )\times BR(h \to XX)\mathcal{L}}\ ,
\label{eqn:sigstr}
\end{equation}
where $\epsilon$ is the selection efficiency, $n^{exp}_{XX}$ is the number of signal events and $\mathcal{L}$ the luminosity.

Since, in MSSM, the Lorentz structure of the Higgs couplings to SM particles remains unchanged, the efficiencies appearing in Eq.(\ref{eqn:sigstr}) are the same as in SM.
This allows us to easily compare the experimentally reported values of $\hmu$ with theoretically calculated values  in the pMSSM. 

While initially both CMS and ATLAS had reported values of $\hmu$ greater than unity with a larger significance, the situation has changed with collection of more data. We summarize the current information on the $h \to \gamma \gamma$ and $h \to ZZ^*$  signal strengths from both ATLAS and CMS in Table~\ref{muh-gg}.

\begin{table}[!h]
\begin{center}
\begin{tabular}{ccc}
\hline \hline \rule[-2ex]{0pt}{5.5ex} Channel & $\hmu$ & Experiment \\ 
\hline \rule[-2ex]{0pt}{5.5ex}  $h \to \gamma \gamma$ & $1.55^{+0.33}_{-0.28}$ & ATLAS \cite{Aad:2013wqa} \\ 
\rule[-2ex]{0pt}{5.5ex}  $h \to \gamma \gamma$ &  $0.77^{+0.27}_{-0.27}$ & CMS \cite{CMS-PAS-HIG-13-005}\\ 
\hline
\rule[-2ex]{0pt}{5.5ex}  $h \to Z Z^*$ &  $1.43^{+0.40}_{-0.35}$ & ATLAS \cite{Aad:2013wqa}\\ 
\rule[-2ex]{0pt}{5.5ex}  $h \to Z Z^*$ &  $0.92^{+0.28}_{-0.28}$ & CMS \cite{CMS-PAS-HIG-13-005}\\ 
\hline \hline 
\end{tabular}
\end{center}
\caption{The present values of $h \to \gamma \gamma$ and $h \to Z Z^*$ signal strengths measured at LHC.}
\label{muh-gg}
\end{table}

It has been recently argued that the presence of light staus can possibly enhance the decay width $\Gamma(h \to \gamma \gamma)$~\cite{Carena:2011aa,Carena:2012gp,*Carena:2012xa}. However, such an enhancement of the width comes at a cost of making the Electroweak vacuum meta-stable, thus restricting the  possible enhancement of the width in the framework of the MSSM~\cite{Kitahara:2012pb,*Carena:2012mw,*Carena:2013iba,*Kitahara:2013lfa}.

From Eq.(\ref{eqn:sigstr}), it is clear that one should study not just the enhancement in partial width but investigate the behavior of the branching ratios (BR) as well, which can be significantly different. Thus one must really check if an excess in the BR can be generated in the MSSM. We study this in the current note.

\begin{figure}
\centering
\includegraphics[width=0.48\textwidth]{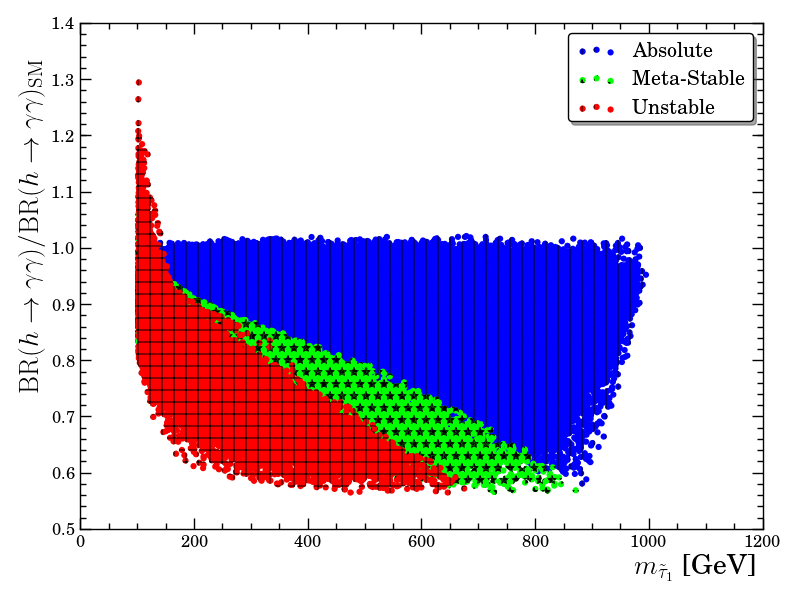}
\includegraphics[width=0.48\textwidth]{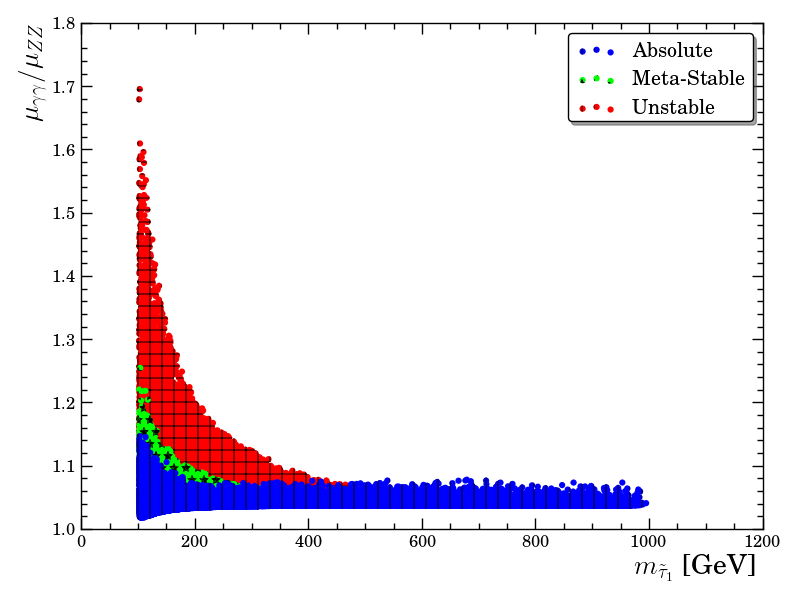}
\caption{Left: Ratio of the branching ratios of $h \to \gamma \gamma$. Right: $\mathcal{R}_{ZZ}(\gamma\gamma)$. The stop mass in the above figures is $\approx 950$ GeV and $\mu$ is set to be positive. Points that correspond to the EWSB vacuum being unstable are given in red (checkered), meta-stable in green (stars) and stable in blue (vertical lines).}
\label{fig:staus}
\end{figure}

The behavior of the BR of the various Higgs decays is controlled by the dominant $h\to b\bar{b}$ BR and hence by the value of the coupling of the Higgs to the bottom quarks. In the MSSM this coupling takes the form~\cite{Carena:1994bv,Djouadi:rep2} 
\begin{align}
g_{hbb}=-\frac{\sin{\alpha}}{\cos{\beta}(1+\Delta_b)}\left( 1 - \frac{\Delta_b}{\tan{\alpha}\tan{\beta}} \right)\ .
\end{align}
Here $\Delta_b$ is the correction from the stop-chargino and sbottom-gluino loops and $\alpha$ corresponds to the
mixing angle between the two CP even Higgses. In the decoupling regime where $\sin\alpha \to -\cos\beta$ an enhancement (suppression) of this coupling depends on whether $|\sin\alpha/\cos\beta|$ is greater (less) than unity. The sign of $\Delta_b$  controls the magnitude of suppression or enhancement of the coupling.
Increasing (decreasing) $g_{hbb}$ results in a reduction (enhancement) in all other branching ratios.
Note also that there is a significant theoretical uncertainty in the determination of this 
BR stemming from scale uncertainties in the value of the bottom quark mass and  $\alpha_s$.
 
In order to detect signs of new physics in the loop induced $h\to \gamma \gamma$ couplings, one should therefore look at quantities that are independent of the branching ratio. Such an observable is the ratio
 \begin{align}
D_{VV}(XX)=\frac{\Gamma(h \to XX)}{\Gamma (h \to VV)}
\end{align}
Where $V$ corresponds to the massive gauge bosons.
 A large amount of the theoretical and experimental uncertainties cancel out when one uses this ratio~\cite{Djouadi:2012rh}.
$D_{VV}(XX)$ is directly related to the partial widths of each of the channels and hence the uncertainty from $h \to b\bar{b}$ is removed.
The dominant contribution to $h\to \gamma \gamma$ comes from the $W$ boson loop. Hence a deviation from unity in $D_{WW}(\gamma\gamma)  $  would signal the presence of additional particles in the loop. Staus would add to the $W$ loop contribution constructively and result in a enhancement of $D^{\gamma\gamma}_{WW}$.

In order to compare with the SM we plot the ratio 
\begin{align}\label{eqn:Rgg}
\mathcal{R}_{VV}(XX) &= \frac{D^{BSM}_{VV}(XX)}{D^{SM}_{VV}(XX)}
= \frac{\Gamma^{BSM}(h \to XX)}{\Gamma^{BSM}(h\to VV)}\frac{\Gamma^{SM}(h\to VV)}{\Gamma^{SM}(h\to XX)} \notag \\
&= \frac{\Gamma^{SM}(h \to VV)}{\Gamma^{BSM}(h\to VV)} \frac{\Gamma^{BSM}(h\to XX)}{\Gamma^{SM}(h\to XX)} \notag \\
&= \frac{\mu_{XX}}{\mu_{VV}}\ .
\end{align}
Here $\mu_{XX}=\frac {\sigma^{MSSM}(pp\to h)\times BR^{MSSM}(h\to XX)}{\sigma^{SM}(pp\to h)\times BR^{SM}(h \to XX)}  $, is different from $\hmu$ defined earlier. This quantity can now be easily compared to the experimental numbers.

As described before for the case of light stops here also we generate the SUSY spectrum using {\tt SuSeFLAV} \cite{Chowdhury:2011zr}, we use {\tt CosmoTransitions} \cite{Wainwright:2011kj} in order to calculate the transition rate of the decay of the false vacuum and we evaluate the Higgs branching ratios using the package {\tt HDECAY} \cite{Djouadi:1997yw}. The parameter ranges we scan in this case is as follows
\begin{align}\label{scan-ranges-stau}
\tan\beta &\in [5,\ 60] \notag \\ 
m_{\tilde{L}_{33}} & \in [100,1000]\ \gev \notag \\
m_{\tilde{E}_{33}} & \in [100,1000]\ \gev \notag \\
\mu &\in [100,1500]\ \gev\ .
\end{align}
We set all the first two generation of sfermion masses to be at $1000$ GeV, the third generation of squarks are set at $m_{\tilde{Q}_{33}} = 1100\ \gev$, $m_{\tilde{U}_{33}} = 1000\ \gev$ and $m_{\tilde{D}_{33}} = 850\ \gev$. To ensure the Higgs sector to be in the decoupling regime we set $M_A=1000$ GeV. The stop trilinear coupling is set to $A_t = 1300$ GeV to ensure the lightest Higgs mass in the LHC observed range while the rest of the trilinear couplings, including $A_{\tau}$, are set to zero. The gaugino masses are set at $M_{1} = 100$ GeV, $M_2  = 300$ GeV and $M_3 = 1000$ GeV.
The scalar potential used, involving only the stau and Higgs fields, is described in appendix~\ref{app:stau}.

In the left plot of Fig.~\ref{fig:staus} we show the variation of the ratio of the branching ratios $(RBR=\frac{BR^{MSSM}(h\to \gamma \gamma)}{BR^{SM}(h \to \gamma \gamma)})$ with the lightest stau mass. 
We see that for our choice of parameters, $RBR$ varies between 0.6 and 1.2. While regions of meta-stability are correlated to the stau mass and as expected decrease with stau mass, there is no clear correlation of these regions with $RBR$. Note also that, as discussed earlier, the values of $RBR$ are highly dependent on the branching ratio of Higgs to bottom quarks. On the other hand when we plot $\mathcal{R}_{ZZ}(\gamma \gamma)$ against the stau mass, we see that as the value of this quantity increases the EWSB vacuum becomes meta-stable and for $\mathcal{R}_{ZZ}(\gamma \gamma) \gtrsim 1.2$ the vacuum becomes unstable. Therefore, not only is there a correlation with the (meta)stability of the vacuum but also an unambiguous comparison with experimentally reported numbers is possible.
We note that the present central values of $\mathcal{R}_{VV}(XX)$ ($1.08$(ATLAS), $0.84$ (CMS)) as can be calculated from Table~\ref{muh-gg} fall in regions that are far from dangerous unstable regions.

\section{Summary and Outlook}
\label{sec:summary}

The detection of the Higgs particle with mass $\sim 125$ GeV implies large values for the trilinear soft terms in MSSM models. It has been known for a while that such large values of these couplings lead to charge and color breaking minima in the scalar potential. In this work we re-derive the constraints that arise from the presence of such minima.
We have chosen a more general framework of MSSM parameters \`a la pMSSM, such that our bounds will have a wider applicability. Our analysis leads us to two major observations. Firstly, our bounds on the low scale MSSM parameters are more stringent than those derived in Ref.~\cite{Kusenko:1996jn}. This has implications for the allowed values of the $X_t$. We find that for $|X_t| \gtrsim \sqrt{6m_{\stopl}m_{\stopr}}$ the electroweak vacuum becomes first meta-stable and then quickly unstable. This value of $X_t$ coincides with the value at which the 1-loop correction to the Higgs mass is maximized.
We also find that in the four field analysis the contribution from $\mu$ starts becoming important when $\mu \sim m_{\stopl,\stopr}$, a feature that was absent in the three field analysis.

Secondly, for the staus, we re-derive the stability bounds and study the implications for Higgs signal strengths. We show that the ratios of the signal strengths provides an unambiguous probe of new physics.

It is now understood that the measured value of the Higgs mass lies in a 
very special (critical) range as far as stability of the Standard Model potential
is considered.  It is surprising to see this feature replicates itself in the MSSM
too, albeit for lightest stop (less than 1 TeV). Further studies are required to 
understand this feature better within the supersymmetric context.

\textbf{Note Added:} As we were finishing our paper, two preprints appeared on the arXiv which deal with the same subject. Ref.~\cite{Camargo-Molina:2013sta} studies the CCB minima confining themselves exclusively to cMSSM models. Ref.~\cite{Blinov:2013uda} presented preliminary results for MSSM with restricted boundary conditions of the type $m_{Q_{3}}^2 = m_{u_3}^2$. They too notice departures from the bounds of Ref.~\cite{Kusenko:1996jn}.

\section*{Acknowledgments}
We would like to thank Alexandre Arbey, Abdelhak Djouadi, Farvah Mahmoudi, Margarete M\"{u}ehlleitner, Michael Spira and Carroll Wainwright for useful inputs and fruitful discussions. We acknowledge important input and discussions with Fabio Zwirner. We would like to thank Bala for carefully reading the manuscript. We thank Miko\l{}aj Misiak for pointing out a few typographical mistakes in the manuscript. The work of DC has been partly supported by the European Research Council under the European Union’s Seventh Framework Programme (FP/2007-2013) / ERC Grant Agreement n. 279972. KAM acknowledges the financial support from CSIR India, the french CMIRA and ENIGMASS Labex and LAPTh, Annecy-le-Vieux for hospitality where part of this work was done. RMG wishes to thank the Department of Science and Technology, Government of India, for support under grant no. SR/S2/JCB-64/2007. SKV thanks Department of Science and Technology, Government of India, for support under grant no. SR/S2/RJN-25/2008.

\appendix
\section{Scalar potential with staus}
\label{app:stau}

The 1-loop scalar potential in MSSM including the up type Higgs ($\hu$), left handed stau ($\staul$) and right handed stau ($\staur$) fields is expressed as below
\begin{align}
\mathcal{V} =& \lb \msq{\hu} + \mu^2 \rb \left|\hu\right|^2 + \msq{\staul} | \staul |^2 + \msq{\staur} \left| \staur \right|^2 - \lb \yuk{\tau} \mu \hu^* \staul \staur + {\rm h.c.} \rb + \yuk{\tau}^2 | \staul \staur |^2 \notag \\
&\ + \frac{g_2^2}{8} \lb |\staul|^2 - |\hu|^2 \rb^2 + \frac{g_1^2}{8} \lb |\staul|^2 - 2 |\staur|^2 - |\hu|^2 \rb^2 + \frac{g_2^2+g_1^2}{8} \delta_H |\hu|^4
\end{align}
where
\begin{align}
\delta_H \approx \frac{3}{\pi^2} \frac{\yuk{t}^4}{g_2^2 + g_1^2} \log \frac{\sqrt{m_{\tilde{t}_1} m_{\tilde{t}_2}}}{m_t}
\end{align}
is the leading term of the full one loop corrected potential arising from top and stop contributions.

\bibliographystyle{apsrev4-1}
\bibliography{vac-stab}

\end{document}